\documentclass[aps,pre,twocolumn,showpacs]{revtex4}
\usepackage{graphicx,color,epsfig}
\usepackage{amssymb}
\begin{document}

\title{Anomalous lifetime distributions and topological traps in ordering dynamics}
\author{X. Castell\'o$^{*,1}$, R. Toivonen$^{*,1,2}$, V. M. Egu\'{\i}luz$^{1}$, J. Saram\"{a}ki$^{2}$, K. Kaski$^{2}$ and M. San Miguel$^{1,2}$}

% \author{J. C. Gonz\'alez-Avella$^1$,  M. G. Cosenza$^1$, and K. Tucci$^{1,2}$}

\affiliation{$^1$Unidad de F\'{\i}sica Interdisciplinar-IMEDEA (CSIC-UIB),
  E07122 Palma de Mallorca, Spain \\
$^2$Laboratory of Computational Engineering, Helsinki University
  of Technology, P.O. Box 9203, 02015 HUT, Finland}

\email{xavi@imedea.uib.es, rtoivone@lce.hut.fi. http://ifisc.uib.es}

\date{\today}


\begin{abstract}
We address the role of community structure
of an interaction network in ordering dynamics, as well as associated
forms of metastability. We consider the voter and AB model dynamics in
a network model which mimics social interactions. The AB model
includes an intermediate state between the two excluding options of
the voter model. For the voter model we find dynamical metastable disordered
states with a characteristic mean lifetime. However, for the AB dynamics
we find a power law distribution of the lifetime of metastable
states, so that the mean lifetime is not representative of the dynamics.
These trapped metastable
states, which can order at all time scales,
originate in the mesoscopic network structure.
\\\\  ** These authors contributed equally to this work.
% \\\\ http://ifisc.uib.es
% \\\\ e-mail: xsavi@imedea.uib.es, rtoivone@lce.hut.fi \\ http://ifisc.uib.es

\end{abstract}

\maketitle

%%%%%%%%%%%%%%%%%%%%%%%%%%%
\section{\label{sec:intro} Introduction} 
Statistical mechanics and complex network theory have been applied
to different disciplines, ranging from biology to sociology. From
this perspective, social systems are modelled as a collection of
agents, located at the nodes of a network, interacting through
simple rules. Social networks of human interaction are structured
into cohesive groups \cite{scott}, and increased knowledge of this
structure \cite{AmaralPNAS,BogunaSocial,nd} has sparked the
creation of new network models
\cite{Jin2001,Davidsen2002,SecederModel,BogunaSocial,WongSpatial,LCEcommNet1}.
These models allow us to study the effect of the structure of
social interactions on the dynamics taking place on the networks,
and on the associated collective phenomena emerging from the
interactions among the agents.

The mesoscopic structure of a social network, and in particular
its community structure, has been found to influence dynamics
taking place on it in ways that cannot be explained by global
level statistics in several cases
\cite{nd,PDInSocialNetworks,lambiotte2006}. In this paper we
address the role of such mesoscopic structure on ordering dynamics
or consensus processes: the question is when the interaction of
agents with several options leads to an ordered state with a
single option (consensus) or when disordered states (possibly
metastable), with coexistent options prevail. We consider two
dynamical models. The first one is the prototype \emph{voter
model} \cite{Holley1975} whose dynamics in complex networks is
known to be generally determined by global properties such as the
effective network dimensionality \cite{voterInNetworks}. Secondly,
we consider the \emph{AB model} \cite{Bilingual1} introduced to
describe language competition, which gives a natural context for
the community concept. These two dynamical models are studied in a
class of networks \cite{LCEcommNet1} incorporating nontrivial
community structure which introduces structural correlations.

%%%%%%%%%%%%%%%%%%%%%%%%%%%%
\section{\label{sec:models}Two dynamical models of competing options}
The \emph{voter model} \cite{Holley1975} concerns the competition
of two equivalent but excluding options A and B. The state of a
node is updated by imitation of a randomly chosen neighbor. The
\emph{AB model} \cite{Bilingual1} includes a third non-excluding
mixed AB state, with the additional rule that a node cannot change
state from A to B or vice versa without going through the AB
state. In studies of dynamics of language competition, the voter
model gives a microscopic version \cite{Stauffer_Castello_2006} of
the Abrams-Strogatz \cite{Abrams_2003} model for the competition
of two socially equivalent languages. In this context the third
state of non-excluding options of the \emph{AB model} is naturally
associated with bilingualism \cite{Wang_2005_TRENDS_Ecology}. More
generally the \emph{AB model} describes competition of two
equivalent social norms which can coexist at the individual level.

In both models, an agent changes its state with a probability
which depends on the states of its neighbors. The fraction of
first neighbors in state A [B, AB] of an agent is
called the \emph{local density} of A, $\sigma_A$ [$\sigma_{B}$,
$\sigma_{AB}$]. For the voter model, the state AB is not allowed
and the probabilities of a node changing state are defined as
follows:
\begin{equation}
\label{eq:voter}
  p_{A\to B} = \sigma_B, \qquad \qquad \;  p_{B\to A} = \sigma_A~.
\end{equation}
The AB model is defined by the following update rules:
\begin{eqnarray}
p_{A\to AB} = \frac{1}{2} \sigma_B, & & \; p_{B\to AB} =
\frac{1}{2} \sigma_A \label{eq:ABa} \\
p_{AB \to A} = \frac{1}{2} (1-\sigma_B), && p_{AB \to B} = \frac{1}{2} ( 1- \sigma_A)~.
\label{eq:ABb}
\end{eqnarray}

In our simulations we start from random initial conditions for the
state of the agents in a network with $N$ nodes (see below) and we
use random asynchronous node update: at each time step a single
node is randomly chosen and updated according to the transition
probabilities Eq.~(\ref{eq:voter}) or
Eqs.~(\ref{eq:ABa})-(\ref{eq:ABb}). We normalize time so that
every unit of time includes $N$ time steps.

A question of interest is under which conditions consensus is
reached (all nodes hold the same option), and which is the process
of emergence and growth of spatial domains where the nodes are in
the same state ({\it coarsening}). Both models are symmetric by
interchange of A and B, so that reaching consensus in either of
these two states is a symmetry breaking process. To describe the
dynamics of the system we use as order parameter the
\emph{interface density} $\rho$, which is defined as the fraction
of links which connect nodes in different states. The ensemble
average interface density $\langle \rho \rangle$ is considered,
where the ensemble average, indicated as $\langle \cdot \rangle$,
denotes average over realizations of the stochastic dynamics
starting from different random initial conditions. Interface
density decreases as domains grow in size. If one of the states
becomes dominant, the interface density decreases along with the
disappearing state. Zero interface density indicates that an
absorbing state, consensus, has been reached. Coarsening in the
voter model is driven by interfacial noise, while for the AB model
earlier results indicate that coarsening is curvature driven:
boundaries tend to straighten out, reducing curvature and leading
to the growth of spatial domains \cite{Bilingual1}. It turns out
that domains of AB agents are never formed. Instead, AB agents
place themselves in the interface between A and B domains.

The dependence of the voter model dynamics on network
dimensionality, disorder and degree distribution has been
carefully studied
\cite{dornicVoter,voterInSmallWorld,Suchecki2005,voterInNetworks}.
A main result is that $d=2$ is the critical dimensionality for
this model. This means that for $d\leq2$ there is coarsening, i.e.
unbounded growth (in the thermodynamic limit) of domains in which
all nodes are in the same state. However for $d>2$ there is no
coarsening beyond an initial transient. In finite networks of
$d>2$ there exist long-lived metastable states in which $\rho$
takes a plateau value. The inverse of this plateau value is the
characteristic size of coexisting A and B domains. Eventually a
finite size fluctuation takes the system to one of the two
consensus absorbing states. We note that complex networks are
typically high dimensional structures for which these metastable
states naturally occur \cite{voterInNetworks}.

Coarsening processes leading to consensus often come to a halt due
to the appearance of metastable states that can be of different
nature. Coarsening and metastable properties depend on the
dynamical model as well as on network characteristics. The type of
metastability encountered for the voter model is characterized by
the fact that all realizations of the process are of the same
class (qualitatively similar) and that the metastable states have
a finite lifetime for a finite system. For the voter model the
mean lifetime of these states scales as $\tau \sim N$
 \cite{voterInNetworks}. We call this type of metastable states
{\it dynamical metastable states}. A different type of
metastability, which we call {\it trapped metastable states},
occurs in situations in which different realizations of the
process are of different type. While some of them follow a
coarsening process until finite size effects come into play,
others get stuck in topological traps. The latter correspond to
trapped metastable states that can be of two types: they might
have a finite lifetime in finite systems, as it occurs for the
\emph{AB model} with stripe-like configurations in regular two
dimensional lattices \cite{Bilingual1}, or they might be
infinitely long lived as it occurs in zero temperature kinetic
Ising models \cite{Spirin01,CastellanoParisi05,CastellanoPastor06,
Boyer_2003}. In summary, different forms of metastability can
appear for the voter and \emph{AB} models considered here, but
every realization is expected to have a finite lifetime in a
finite system.

%%%%%%%%%%%%%%%%%%%%%%%%%%%%%%
\section{\label{sec:networks} A class of social type networks}
Several models have been designed to capture some of the
characteristics of social networks, based on mechanisms such as
geographical proximity \cite{WongSpatial}, social similarity
\cite{SecederModel,BogunaSocial}, and local search
\cite{Jin2001,Davidsen2002,LCEcommNet1}. A combination of random
attachment with local search for new contacts has proved fruitful
in generating cohesive structures as well as well-known features
of social networks, such as assortativity, broad degree
distributions, and community structure \cite{LCEcommNet1}. The
term ``community'' is typically used in the context of groups of
nodes with dense internal and sparse external connections; exact
definitions differ
\cite{GirvanPNAS,NewmanCommunity,OverlappingCommunities,GuimeraCartography}.
The community structure leads naturally to high values of the
clustering coefficient and to positive degree-degree correlations.

\begin{figure}[tb]
\centering
\includegraphics[height=0.25\linewidth]{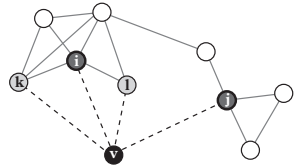}
 \caption{Growth process of the network. The new vertex
 $v$ links to one or more randomly chosen initial contacts
 (here $i,j$) and possibly to some of their neighbors (here $k,l$).
}
 \label{fig:schematic}
\end{figure}

The algorithm to generate this class of networks consists of two
growth processes: 1) random attachment, and 2) implicit
preferential attachment resulting from following edges from the
randomly chosen initial contacts. The local nature of the second
process gives rise to high clustering, assortativity and community
structure. Starting from any small connected seed network of
$N_{0}$ vertices, new nodes are added as follows (see
Fig.~\ref{fig:schematic}): i) Pick $n_{init}\geq1$ random nodes as
initial contacts. ii) Pick $n_{sec}\geq0$ neighbors of each
initial contact as secondary contacts. iii) Connect the new node
to the initial and secondary contacts.

%%%%%%%%%%%%%%%%%%%%%%%%%%%%%%%%%%%%%%%%%%%%%%%%%%%%%%%%%Fig. 1

Throughout this paper, we will use the {\it standard parameters}
\cite{LCEcommNet1}: the number of initial contacts is selected
according to the probabilities $p(n_{init}=1)=0.95$,
$p(n_{init}=2)=0.05$; and the number of secondary contacts from
each initial contact, $n_{sec}$, is chosen from a uniform
probability distribution between 0 and 3; the initial seed
contains $N_0 =10$ nodes.

\begin{figure}[tb]
\centering
\includegraphics[width=0.40\linewidth, angle=0,scale=1.0]{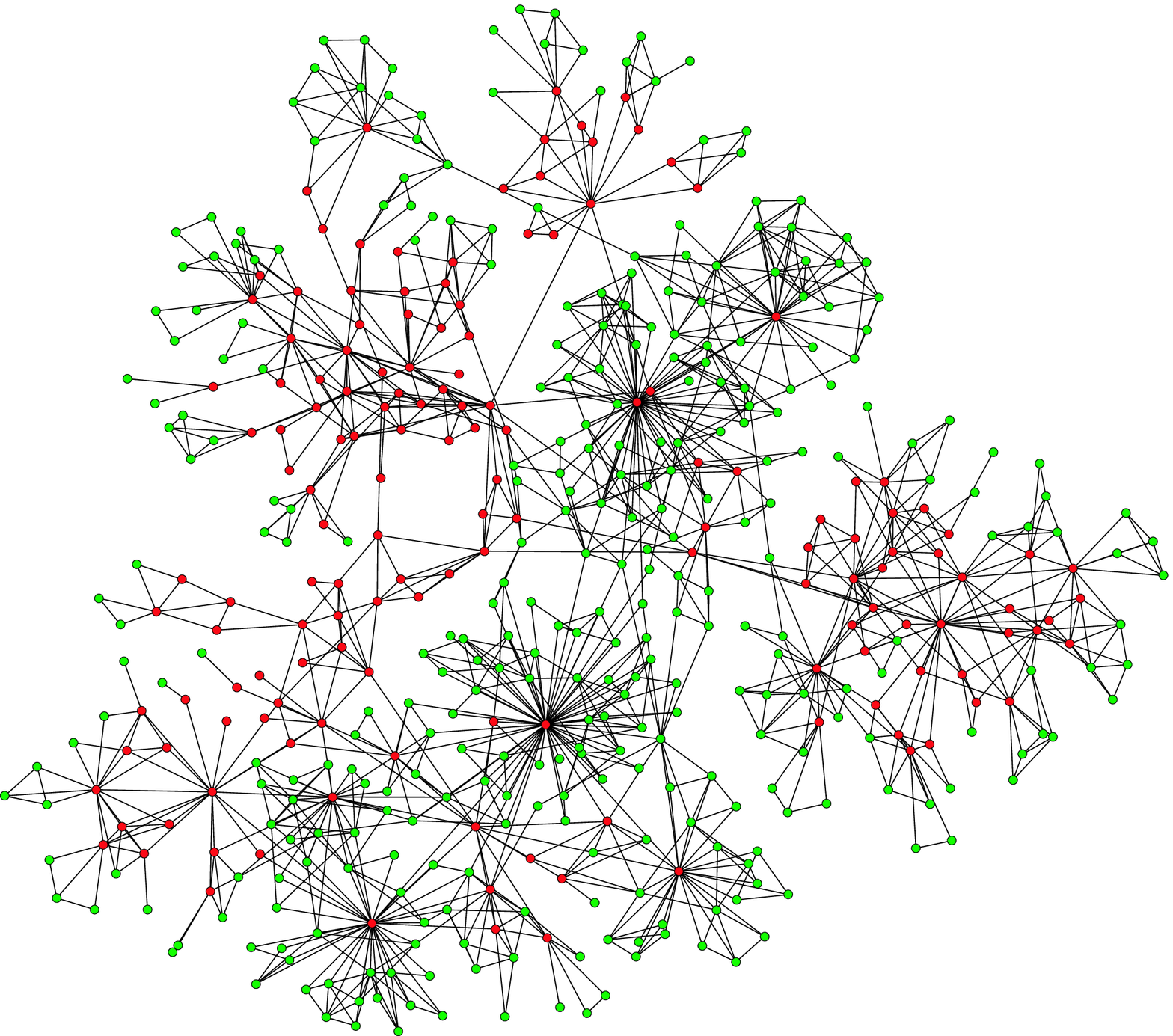}
\includegraphics[width=0.55\linewidth, angle=0,scale=1.0]{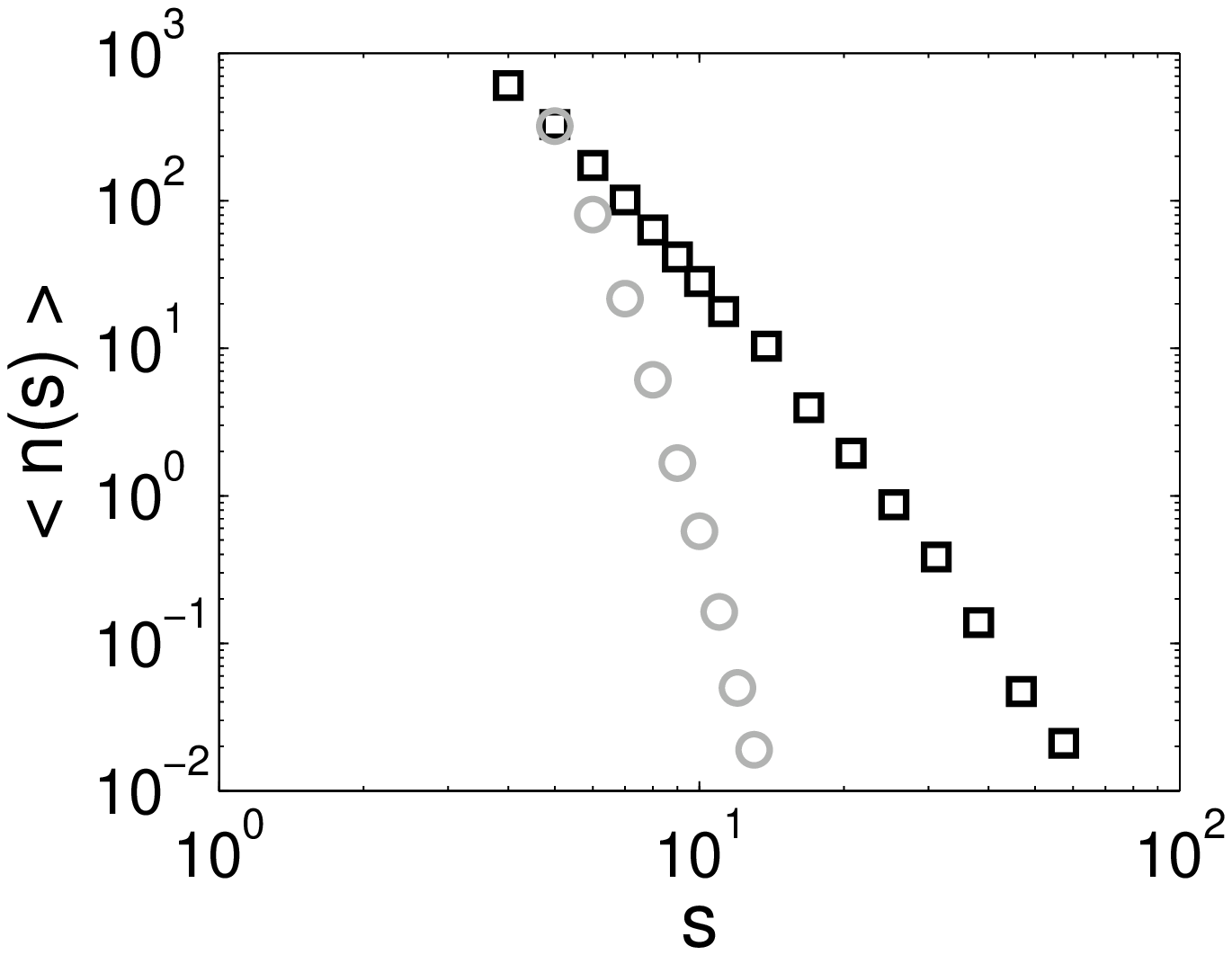}
\caption{Left: A partial view of the network centered on a randomized
selected node. Right: Average number $\langle n(s) \rangle$ of
$k$-clique-communities of size $s$ for $k=4$ ({$\square$})
and $k=5$ ({$\bigcirc$}), in networks
of size $N=10\,000$, averaged over 400 realizations.}
\label{fig:snowball}
\end{figure}

The degree distributions of the resulting networks are found to decay slower
than exponential \cite{LCEcommNet1}. Using the $k$-clique algorithm
\cite{OverlappingCommunities} for detecting communities, a broad
distribution of community sizes is found in the model
(Fig.~\ref{fig:snowball}).

For reference, we use randomized versions of the same networks,
where the degree sequence is kept intact but edges are randomly
rewired under the restriction that the network must stay connected
\cite{Maslou}. This eliminates community structure, clustering,
and degree correlations. The randomized networks are therefore
locally treelike with very few loops.

%%%%%%%%%%%%%%%%%%%%%%%%%%%%%%%%%%%%%%%%%%%%%%%%%%%%%%%%%%%%Fig. 2

%%%%%%%%%%%%%%%%%%%%%%%%%%%%%%%%%%%%%%
\section{\label{sec:findings} Results}
We have considered the update rules Eqs.~(\ref{eq:voter}) for the
voter model, or Eqs.~(\ref{eq:ABa})-(\ref{eq:ABb}) for the AB
model in the class of networks described above. We followed the
development over time of the interface density and of the fraction
of runs that had not yet reached consensus at any particular time.
When results for the original and randomized networks differ, we
can conclude that structural characteristics other than the degree
distribution are responsible for the differences.

\subsection{\label{sec:interfacedensity} Interface density}
\begin{figure}[tb]
\vspace{0.2cm}
\centerline{\includegraphics[width=0.95\linewidth]{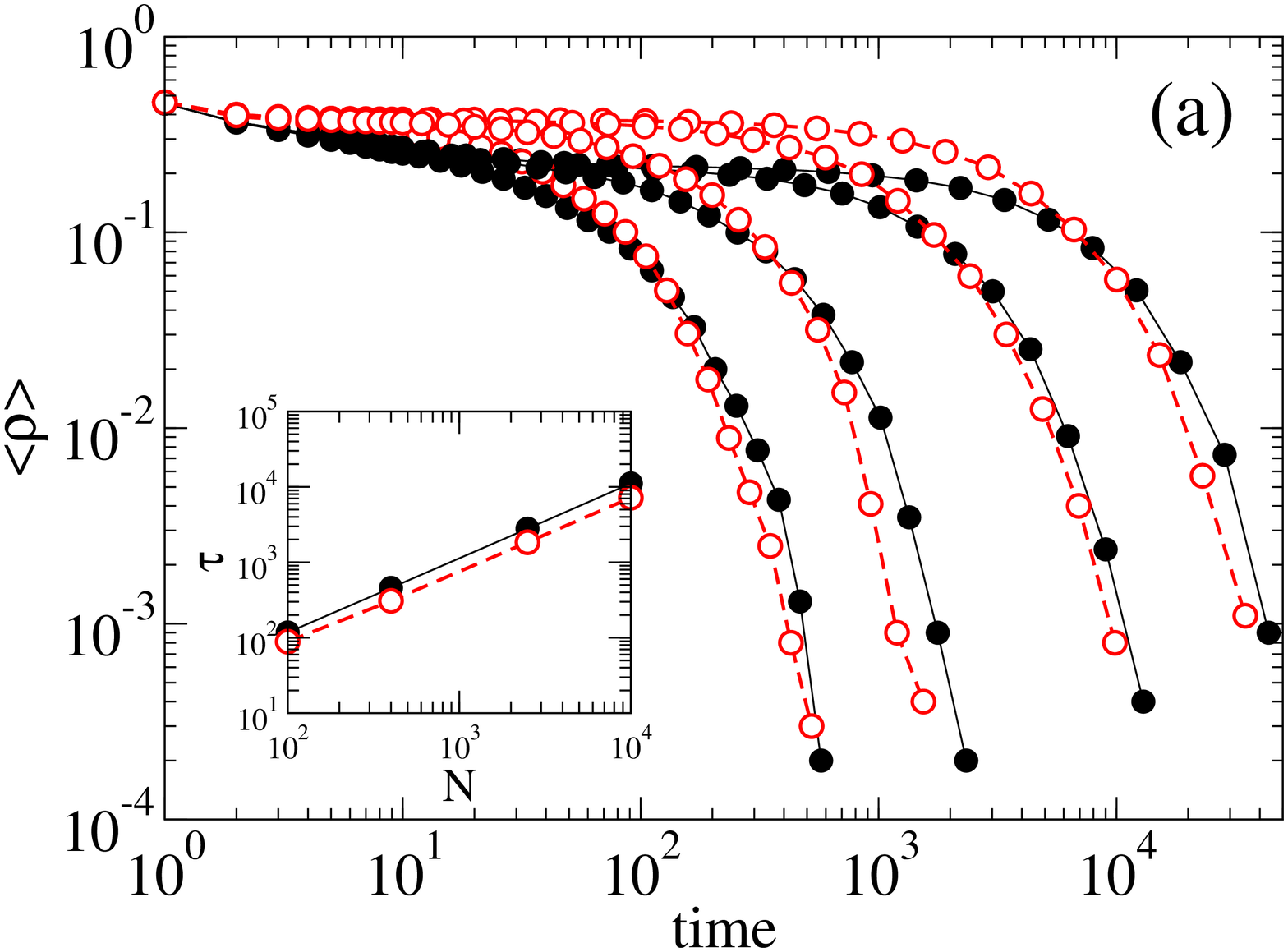}}
\vspace{0.8cm}
\centerline{\includegraphics[width=0.95\linewidth]{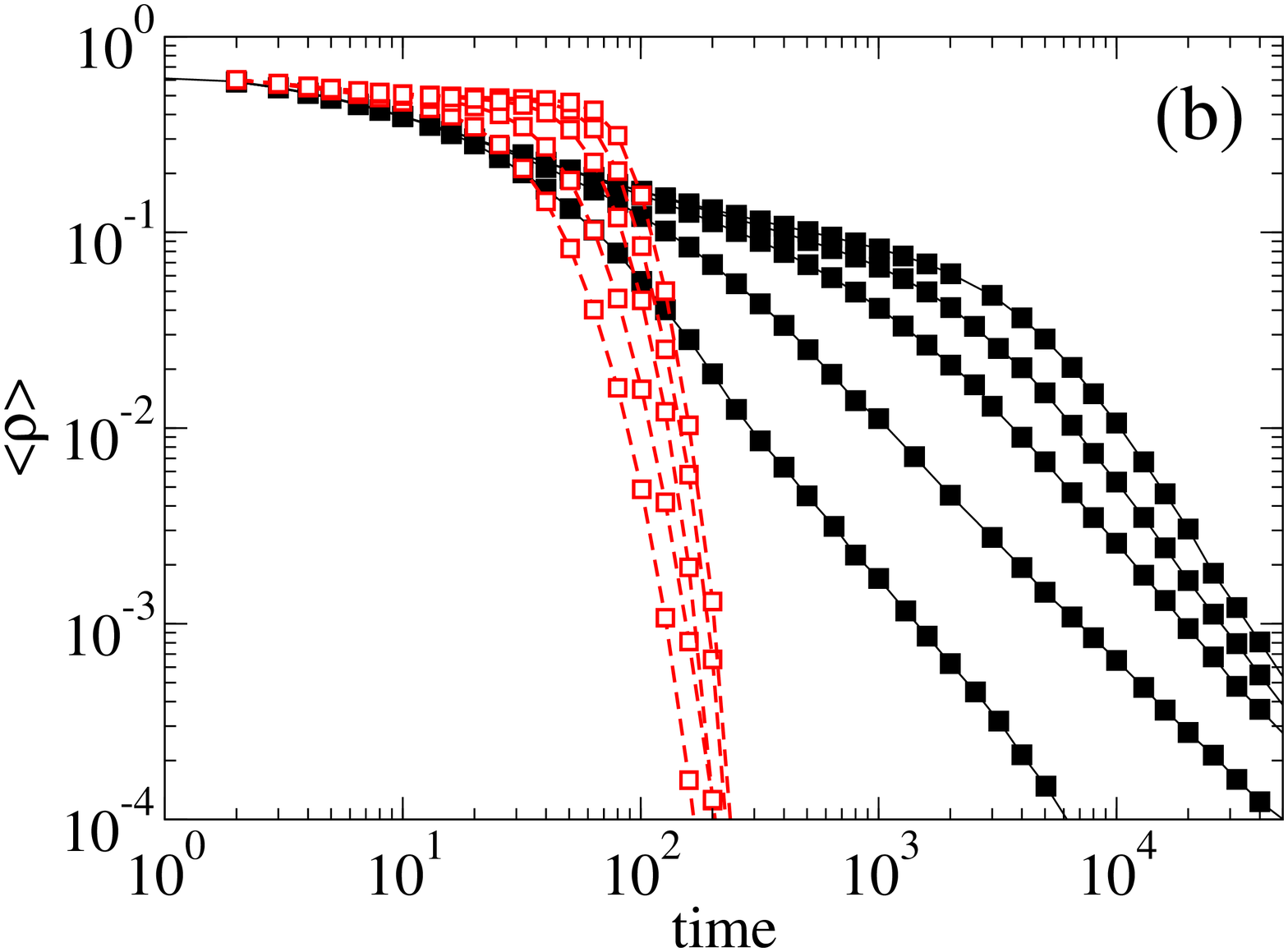}}
\caption{Time evolution of the average interface density
in networks with communities (solid symbols) and randomized networks
(empty symbols) with the same degree sequences. (a) Voter model.
Network sizes increase from left to right: $N=100$, $400$, $2500$,
$10000$. Averages are taken over $100$ different realizations
of the model network, with $10$ runs in each. Inset: time to
reach consensus scales with network size as $\tau\sim N^{\gamma},
\, \gamma \approx 0.96$
for the randomized and $\gamma \approx 0.98$ for the
original networks. (b) AB model. Network sizes increase from left
to right: $N=100$, $400$, $2500$, $10000$, $40000$. Averages taken
over $400-5000$ realizations (depending on system size) of the model
network, and with $10$ runs in each.}
\label{fig:interfaceDensity}
\end{figure}

The average interface density $\langle \rho\rangle $ on the class
of networks considered here, and on their randomized counterparts is
shown in Fig.~\ref{fig:interfaceDensity}. For the voter model
(Fig.~\ref{fig:interfaceDensity}a), we obtain that the structure
of the network does not alter the qualitative behavior. In both
classes of networks we observe plateau values of $\langle
\rho\rangle $ associated with dynamical metastable states. Still,
the plateau value for networks with community structure is lower
than for the randomized networks, indicating that the typical size
of spatial domains where agents are in the same state is larger.
We also observe in both cases that finite size fluctuations drive
the system to an absorbing state. The characteristic time to reach
consensus (mean lifetime of the metastable state) depends on
network size but it does not depend sensitively on network
structure. The inset in Fig.~\ref{fig:interfaceDensity}a shows that
the time to reach consensus depends linearly on network size for networks
with communities and their randomized counterparts\footnote{The slight
deviation from linear scaling is due to violation of conservation
laws when using node update dynamics on networks with nodes of
very different degree (see \cite{Suchecki2005}).}. These results
support the earlier finding made on networks without mesoscopic
structure that effective dimensionality dominates voter model
behavior \cite{voterInNetworks}.

%%%%%%%%%%%%%%%%%%%%%%%%%%%%%%%%%%%%%%%%%%%%%%Fig. 3

Figure~\ref{fig:interfaceDensity}b shows the average interface
density for the AB dynamics. We observe significant differences
between the original and the randomized version networks: a
plateau value of $\langle \rho\rangle $ is observed for randomized
networks, while a first dynamical stage of coarsening where spatial
domains grow in size is found for large networks with communities.
The plateau observed in randomized networks indicates that a
dynamical metastable state of the class found in the voter model
for both types of networks is rapidly reached. Moreover, in the
randomized networks there is a fast decay towards an absorbing
state with a characteristic time to reach consensus almost
independent of system size. For the networks with a community
structure we observe two dynamical stages in the evolution of
$\langle \rho \rangle$. After an initial power law associated with
coarsening there appears a second power law tail in the approach
to the absorbing state. This last power law decay indicates that
the mean lifetime to reach consensus for the AB model does not
characterize the dynamics on these networks and that metastable
states exist at all time scales, as we discuss below. Additionally,
the difference with the randomized networks in several orders of
magnitude for the extinction times, which increases with system
size, shows that the network with communities slows down the
dynamics significantly. All together these results manifest a
sensitivity of the AB dynamics to the mesoscopic network structure
which is not found for the voter dynamics.

\subsection{\label{sec:aliveruns}Fraction of alive runs}
Figure~\ref{fig:aliveruns} shows the fraction $P(t)$ of
realizations still alive at time {\it t}, i.e. the fraction of
realizations which have not reached the absorbing state. For the
voter model, the fraction of alive runs decreases exponentially in
both the original and randomized networks
(Fig.~\ref{fig:aliveruns}-inset), in agreement with previous
results for the voter model in high dimensional complex networks
\cite{voterInNetworks}. A rather different result is found for the
AB model (Fig.~\ref{fig:aliveruns}). In our class of networks, we
find a power law behavior $P(t) \sim t^{-\alpha}, \, \alpha
\approx 1.3 $, so that a mean lifetime of the realizations of the
AB dynamics does not give a characteristic time scale. At any time
there are live realizations which have not reached the absorbing
state. Different parametrizations of the network model (not shown)
produce the same qualitative phenomenon: we have modified the
number of secondary contacts from each initial contact, $n_{sec}$,
using uniform probability distributions between 0 and ${1,2,4}$,
obtaining also a power law of the distribution of alive runs with
an exponent smaller than 2, which indicates the robustness of this
result. This behavior is different from the usual exponential
decay of the tails of $P(t)$ observed for the voter, and AB
dynamics either in regular, small world \cite{Bilingual1}, random
or Barab{\'a}si-Albert scale-free networks (not shown), and
reflects the existence of metastable states at all time scales.
This fact indicates that the anomalous lifetime distribution is
linked to the structure of the network at a mesoscopic level. Such
structure seems to give rise to a number of traps that cause
trapped metastable states at all time scales. To substantiate this
claim we next look at some detailed dynamics.

\subsection{\label{sec:discussion}Discussion}
Further understanding of the dynamical process can be obtained by
considering the measure called overlap, $O$ \cite{nd}. This
characteristic of a link between two nodes tells us essentially
which fraction of their neighbors is shared by the nodes. Within a
community, nodes tend to share many neighbors, and thus overlap is
high, while edges between communities will have low or zero
overlap. Considering dynamics of competing options on a network,
the overlap can be used to identify spatially homogenous domains in
the network: if the average overlap $\langle O \rangle$ of the links in the
interface between domains is low, we may assume that the domain
boundaries follow the community boundaries. On the other hand, if
the overlap at the interfaces is high, it indicates that nodes
within communities are in different states. For the voter model
dynamics we have found that the average overlap of interface links
drops to about 80 percent of the average value $\langle O
\rangle=0.27$ of the whole network, while in the AB model it drops
to under 70 percent. This indicates that in both models the
interfaces between domains lie preferably in low overlap links, so
that domains of the same option follow the community structure,
but in the AB model these domains are correlated with the
communities closer.

%%%%%%%%%%%%%%%%%%%%%%%%%%%%%%%%%%%%%%%%%%%Fig. 4
\begin{figure}[tb]
\vspace{0.2cm}
\centering
\includegraphics[width=0.95\linewidth]{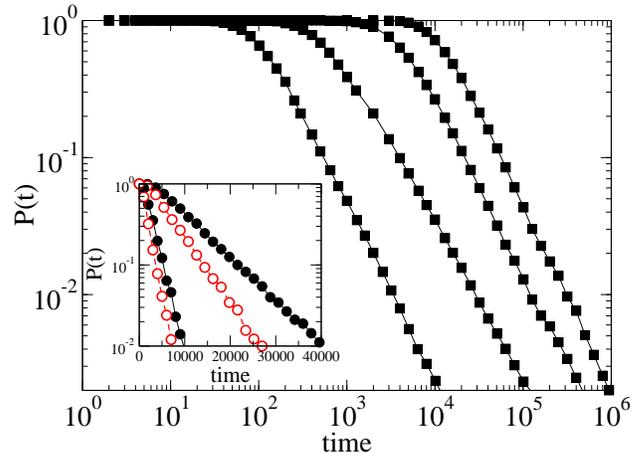}
\caption{Fraction of alive runs in time for networks
with communities (solid symbols) and randomized networks (empty symbols).
AB model (double logarithmic plot); system sizes $N= 100$,
$400$, $2500$, $10000$ from left to right, with averages taken over
different realizations of the network ($400-5000$ depending
on system size), with $10$ runs in each. Inset: voter model
(semilogarithmic plot). System sizes
$N= 2500$, $10000$. Averages are taken over $100$ different
realizations of the networks, with $10$ runs in each.}
\label{fig:aliveruns}
\end{figure}

\begin{figure}[tb]
\centering
\includegraphics[width=0.95\linewidth]{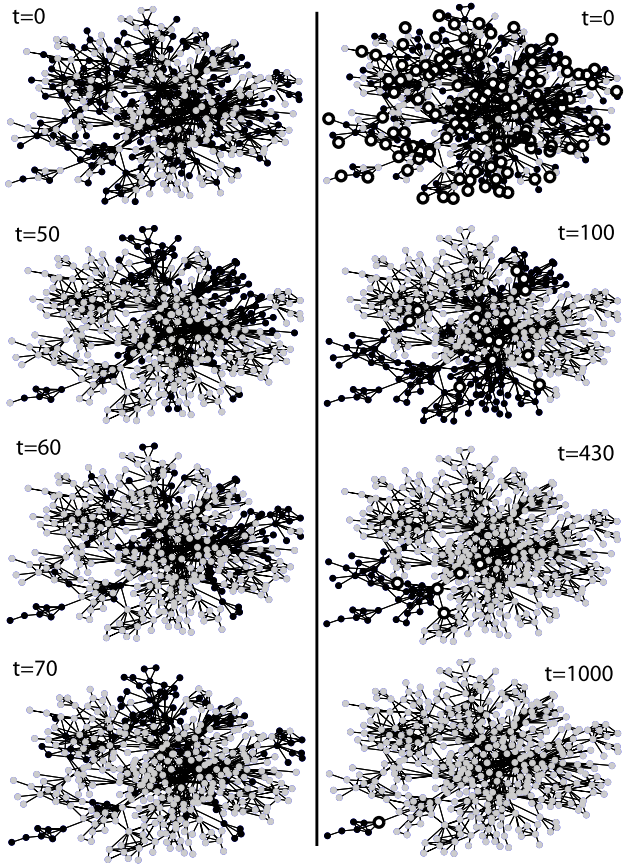}
\caption{Snapshots of the dynamics, with nodes in state A in black, B in grey,
and AB in white circled in black. Simulations start from random initial conditions.
Left: voter model. Right: AB model.}
 \label{fig:snapshots}
\end{figure}

The difference between the two dynamics is better understood by
looking at snapshots of the dynamics (Fig.~\ref{fig:snapshots})
which show the characteristic behavior for each of the models,
starting from random initial conditions ($t=0$). In the voter
model (left) the homogeneous domains of nodes with the same option
appear to follow the community structure, but a particular
community (topological region) may change the option adopted by
the community rather quickly ($t=50,60,70$). At variance with this
behavior, in the AB model (right) spatial domains grow and
homogenize steadily in a community without much fluctuation. For
this dynamics, communities that have adopted a given option, and
which are poorly linked to the rest of the network, take a long
time to be invaded by a different option, acting therefore as
topological traps. As an example of this we show two long lived
trapped metastable state at $t=430$ and $t=1000$, where the
interface stayed relatively stable for a prolonged period ($\sim
100$ and $\sim 1000$ time steps, respectively). These different
behaviors reflect in the community structure two different
interfacial dynamics: interfacial noise driven dynamics for the
voter model, and curvature driven dynamics for the AB model with
agents in the AB state at the interfaces.

%%%%%%%%%%%%%%%%%%%%%%%%%%%%%%%%%%%%%%%%%%%%%%%%%%%%%%%%%%%Fig. 5

%%%%%%%%%%%%%%%%%%%%%%%%%%%%%%%%%%%%%%%%%%%%%%%%%%%%%%%%Fig. 6
\begin{figure}[ht]
\vspace{0.1cm}
\centerline{\includegraphics[width=0.95\linewidth]{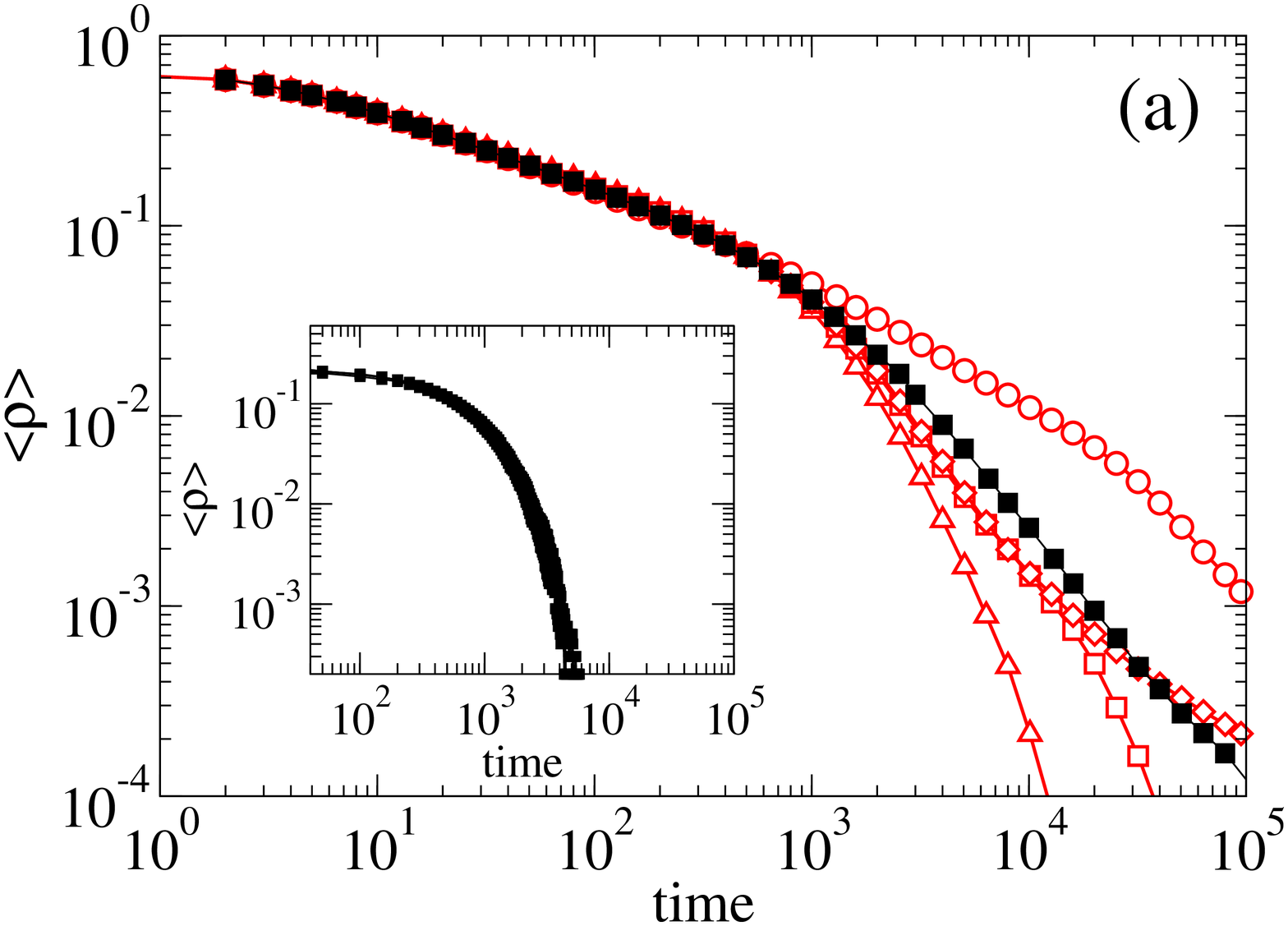}}
\vspace{.8cm}
\centerline{\includegraphics[width=0.95\linewidth]{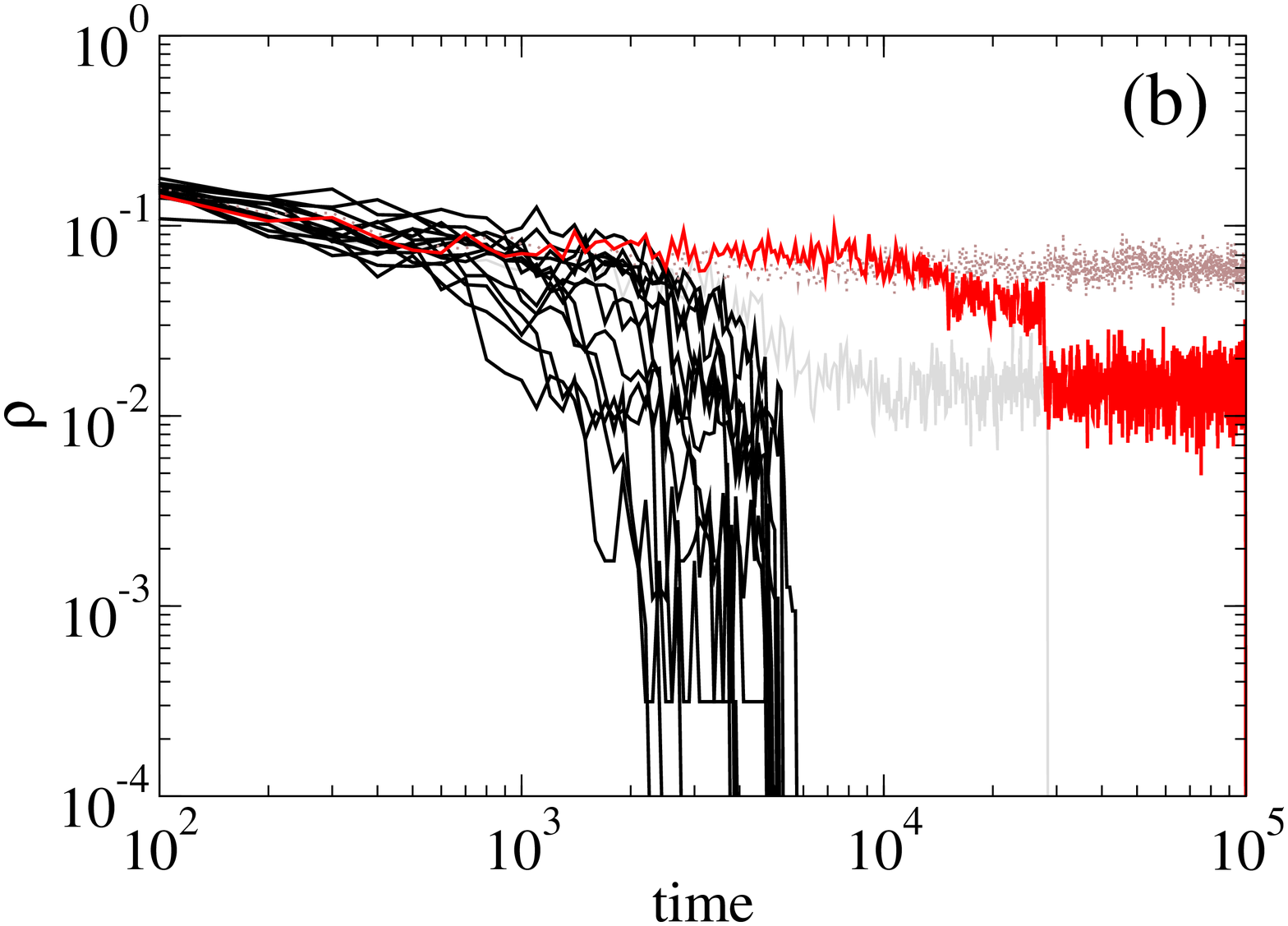}}
\caption{(a) Time evolution for the AB model of the average
interface density on different realizations of the network with
2500 agents; 20000 runs on each (empty symbols). The extreme cases
were selected as examples of networks where {\it trapped
metastable states} (see text) are found {\it often}
(\textcolor{red}{$\bigcirc$}); and found {\it rarely}
(\textcolor{red}{$\triangle$}). For comparison, the average over
500 networks (10 runs on each) is also shown ($\blacksquare$).
Inset: time evolution for the voter model of the average interface
density for four realizations of the networks of 2500 agents; 5000
runs on each network. (b) Time evolution of the interface density
in single realizations of the AB dynamics on a network with 2500
agents. A class of realizations decay to the absorbing state after
a coarsening stage (solid black lines), while others fall in long
lived trapped metastable states. The latter display several
plateaus, indicating hierarchical levels of ordering before
reaching the absorbing state, or cascading between several trapped
metastable states. } \label{fig:N50_greenthing}
\end{figure}

Different realizations of the algorithm to construct the social
type network produce different detailed structures of the network.
The power-law for the fraction of alive runs in
Fig.~\ref{fig:aliveruns} is a statistical effect of the average
over such realizations. The time evolution of the average
interface density on single realizations of the network, $\langle
\rho \rangle $, is shown for the AB dynamics in
Fig.~\ref{fig:N50_greenthing}a. We observe different behaviors in
the second stage of the decay of $\langle \rho\rangle $ depending
on the specific realization of the network: from broad tails to
exponential-like decays, with an intermediate behavior. On the
other hand, and in agreement with our previous discussion, the
voter model dynamics (Fig.~\ref{fig:N50_greenthing}a, inset) is
not sensitive to the details of the network structure. For the AB
model some realizations of the network topology produce
particularly long lived metastable states, while in others,
corresponding to exponential-like decay of $\langle \rho\rangle $,
they are observed rarely. Plots of the interface density of
individual runs on a given network show a class of realizations
with different plateaus (ordering levels) where the system gets
trapped for a long time (Fig.~\ref{fig:N50_greenthing}b). These
trapped metastable states, analogous to those displayed in
Fig.~\ref{fig:snapshots}-right, correspond to the structure in the
network. The variety of traps and associated different lifetimes
seems to be the mechanism that causes an anomalous power law
distribution for the lifetimes.

We note that although the details of each network realization
matter for the occurrence of trapped metastable states, the
community size distribution detected by the
\emph{k}-clique-percolation method \cite{OverlappingCommunities}
is the same for all the network realizations that we have
considered. This and other available statistical methods seem not
to be sufficient to discern between the network topologies
producing many or few trapped metastable states.

%%%%%%%%%%%%%%%%%%%%%%%%%%%%%%%%%%%%%
\section{Summary and conclusions} We have considered two dynamical
models, the voter and the AB model, in order to study metastable
states and the role of community structure in the dynamics of
consensus processes. The voter model dynamics, driven by
interfacial noise, is not particularly sensitive to the mesoscopic
structure of the network: we find that all realizations of the
dynamics are of the same class, leading to a type of dynamical
metastable states shared by other complex networks of high
dimensionality without degree correlations. On the contrary, for
the AB dynamics we find different classes of realizations leading
to a power law distribution for the times to reach consensus. This
is explained in terms of trapped metastable states associated with
the structure of the network. Our result implies that a mean
lifetime for these states does not give a characteristic time
scale of the ordering dynamics. We note that a mean lifetime does
not exist for the zero-temperature kinetic Ising model dynamics on
regular or complex networks \cite{CastellanoPastor06}, due to
realizations that lead to trapped metastable states of infinite
lifetime in finite systems. The novelty of our finding is that we
have realizations with any lifetime. For the AB model in a regular
2D lattice trapped metastable states with stripe-like
configuration have been found \cite{Bilingual1}, but in that case
the distribution of lifetimes is exponential: $P(t) \sim {\mathtt
e}^{-\alpha t}$ and the mean lifetime is representative of the
dynamics. The power-law distribution for the lifetimes originates
here in the multiplicity of different traps that reflects the
mesoscopic structure of the networks. Simpler configurations of
community structure should be considered in the future in order to
gain a deeper understanding of the microscopic mechanisms
underlying consensus dynamics.
% \vspace{15cm}
%%%%%%%%%%%%%%%%%%%%%%%%%%%%%%%%%%%%
\section*{ACKNOWLEDGEMENTS} 
% \acknowledgements 
This work was supported in part by the Finnish
Academy of Science and Letters, Vilho, Yrj\"{o} and Kalle
V\"{a}is\"{a}l\"{a} Foundation, Cost Action P10
(COST-STSM-P10-02707), as well as the Academy of Finland, Center
of Excellence program 2006-2011. We acknowledge financial support
form the MEC (Spain) through project CONOCE2 (FIS2004-00953). X.C.
also acknowledges financial support from a Ph.D. fellowship of the
Govern de les Illes Balears (Spain).


\begin{thebibliography}{10}

\bibitem{scott} Scott J., {\it Social network analysis: A Handbook}, 2nd ed.; Sage, London (2000).
% \bibitem{scott}
%    \Name{Scott J.}
%    \Book{Social network analysis: A Handbook, 2nd ed.}
%    \Publ{Sage, London}
%    \Year{2000}.

\bibitem {AmaralPNAS}Amaral L. A. N.,  Scala A.,  Barth\'{e}l\'{e}my M. and Stanley H. E., Proc. Natl. Acad. Sci. USA {\bf 97} (2000) 11149.
% \bibitem{AmaralPNAS}
%    \Name{Amaral L. A. N.,  Scala A.,  Barth\'{e}l\'{e}my M. \and Stanley H. E.}
%    \REVIEW{Proc. Natl. Acad. Sci. USA}{97}{2000}{11149}.

\bibitem{BogunaSocial} Bogu$\tilde{\textrm{n}}$a M., Pastor-Satorras R., D\'{\i}az-Guilera A. and Arenas A., Phys. Rev. E {\bf 70} (2004) 056122.
% \bibitem{BogunaSocial}
%   \Name{Bogu$\tilde{\textrm{n}}$a M., Pastor-Satorras R., D\'{\i}az-Guilera A. \and Arenas A.}
%    \REVIEW{Phys. Rev. E}{70}{2004}{056122}.

\bibitem{nd} Onnela J.-P., Saram\"{a}ki J., Hyv\"{o}nen  J., Szab\'o G.,  Lazer D., Kaski K., Kert\'esz J. and  Barab\'asi A.-L., Proc. Natl. Acad. Sci. USA {\bf 104} (2007) 7332.

% \bibitem{nd} 
%  \Name{Onnela J.-P., Saram\"{a}ki J., Hyv\"{o}nen  J., Szab\'o G.,  Lazer D., Kaski K., Kert\'esz J. \and  Barab\'asi A.-L.}
%  \REVIEW{Proc. Natl. Acad. Sci. USA}{104}{2007}{7332}.

\bibitem{Jin2001} Jin E. M., Girvan M. and  Newman M. E. J., Phys. Rev. E {\bf 64} (2001) 046132.
 

% \bibitem{Jin2001}
%  \Name{Jin E. M., Girvan M. \and  Newman M. E. J.}
%  \REVIEW{Phys. Rev. E}{64}{2001}{046132}.

\bibitem{Davidsen2002} Davidsen J., Ebel H. and Bornholdt S., Phys. Rev. Lett. {\bf 88} (2002) 128701.

% \bibitem{Davidsen2002}
%  \Name{Davidsen J., Ebel H. \and Bornholdt S.}
%  \REVIEW{Phys. Rev. Lett.}{88}{2002}{128701}.

\bibitem{SecederModel} Gr\"onlund A. and Holme P., Phys. Rev. E {\bf 70} (2004) 036108.

% \bibitem{SecederModel}
%  \Name{Gr\"onlund A.\and Holme P.}
%  \REVIEW{Phys. Rev. E}{70}{2004}{036108}.

\bibitem{WongSpatial} Wong L. H., Pattison P. and Robins G., Physica A {\bf 360} (2006) 99.

% \bibitem{WongSpatial}
%  \Name{Wong L. H., Pattison P.\and Robins G.}
%  \REVIEW{Physica A}{360}{2006}{99}.

\bibitem{LCEcommNet1} Toivonen R., Onnela J.-P., Saram\"{a}ki J., Hyv\"{o}nen J., Kert\'{e}sz J. and Kaski K., Physica A {\bf 371} (2006) 851.
% \bibitem{LCEcommNet1}
% \Name{Toivonen R., Onnela J.-P., Saram\"{a}ki J., Hyv\"{o}nen J.,
% Kert\'{e}sz J. \and Kaski K.} \REVIEW{Physica A}{371}{2006}.

\bibitem{PDInSocialNetworks} Lozano S., Arenas A. and S\'anchez A., {\it preprint} www.arXiv.org/physics/0612124 (2006).
% \bibitem{PDInSocialNetworks}
% \Name{Lozano S., Arenas A. \and S\'anchez A.} \REVIEW{preprint
% www.arXiv.org/physics/0612124}{}{2006}{}.


\bibitem{lambiotte2006} Lambiotte R., Ausloos M. and Holyst J., Phys. Rev. E {\bf 75} (2007) 030101(R).
% \bibitem{lambiotte2006}
%   \Name{Lambiotte R., Ausloos M. \and Holyst J.}
%  \REVIEW{Phys. Rev. E}{75}{2007}{030101(R)}.

\bibitem{Holley1975} Holley R. and Liggett T., Ann. Probab. {\bf 4} (1975) 195.
% \bibitem{Holley1975}
%  \Name{Holley R. \and Liggett T.}
%  \REVIEW{Ann. Probab.}{4}{1975}{195}.

\bibitem{voterInNetworks} Suchecki K., Egu\'{\i}luz V.M. and San Miguel M., Phys. Rev. E {\bf 72} (2005) 036132.
% \bibitem{voterInNetworks}
%  \Name{Suchecki K., Egu\'{\i}luz V.M. \and San Miguel M.}
%  \REVIEW{Phys. Rev. E}{72}{2005}{036132}.

\bibitem{Bilingual1} Castell\'o X., Egu\'{\i}luz V.M. and San Miguel M., New J. Phys. {\bf 8} (2006) 308.
% \bibitem{Bilingual1}
% \Name{Castell\'o X., Egu\'{\i}luz V.M. \and San Miguel M.}
% \REVIEW{New J. Phys.}{8}{2006}{308}.

\bibitem{Stauffer_Castello_2006} Stauffer D., Castell\'o X., Egu\'{\i}luz V.M. and San Miguel M., Physica A, {\bf 374} (2007) 835.
% \bibitem{Stauffer_Castello_2006}
%   \Name{Stauffer D., Castell\'o X., Egu\'{\i}luz V.M. \and San Miguel M.}
%   \REVIEW{Physica A}{374}{2007}{835}.

\bibitem{Abrams_2003} Abrams D. M. and Strogatz S. H., Nature {\bf 424} (2003) 900.
% \bibitem{Abrams_2003}
%   \Name{Abrams D. M. \and Strogatz S. H.}
%   \REVIEW{Nature}{424}{2003}{900}.

\bibitem{Wang_2005_TRENDS_Ecology} Wang W. S.-Y. and Minett J. W., Trends Ecol. Evol. {\bf 20} (2005) 263.
% \bibitem{Wang_2005_TRENDS_Ecology}
%   \Name{Wang W. S.-Y. \and Minett J. W.}
%   \REVIEW{Trends Ecol. Evol.}{20}{2005}{263}.

\bibitem{dornicVoter} Dornic I., Chat\'e H., Chav\'e J. and Hinrichsen H., Phys. Rev. Lett. {\bf 87} (2001) 045701.
% \bibitem{dornicVoter}
% \Name{Dornic I., Chat\'e H., Chav\'e J. \and Hinrichsen H.}
% \REVIEW{Phys. Rev. Lett.}{87}{2001}{045701}.

\bibitem{voterInSmallWorld} Castellano C., Vilone D. and Vespignani A., Europhys. Lett. {\bf 63} (2003) 153.
% \bibitem{voterInSmallWorld}
% \Name{Castellano C., Vilone D. \and Vespignani A.}
% \REVIEW{Europhys. Lett.}{63}{2003}{153}.

\bibitem{Suchecki2005} Suchecki K., Egu\'{\i}luz V. M. and San Miguel M., Europhys. Lett. {\bf 69} (2005) 228.
% \bibitem{Suchecki2005}
%  \Name{Suchecki K., Egu\'{\i}luz V. M. \and San Miguel M.}
%  \REVIEW{Europhys. Lett.}{69}{2005}{228}.

\bibitem{Spirin01} Spirin V, Krapivsky P.L. and Redner S., Phys. Rev. E {\bf 65} (2001) 0116119.
% \bibitem{Spirin01}
% \Name{Spirin V, Krapivsky P.L. \and Redner S.} \REVIEW{Phys. Rev.
% E}{65}{2001}{0116119}.

\bibitem{CastellanoParisi05} Castellano C., Loreto V., Barrat A., Cecconi F. and Parisi
D., Phys. Rev. E {\bf 71} (2005) 066107.
% \bibitem{CastellanoParisi05}
% \Name{Castellano C., Loreto V., Barrat A., Cecconi F. \and Parisi
% D.} \REVIEW{Phys. Rev. E}{71}{2005}{066107}.

\bibitem{CastellanoPastor06} Castellano C. and Pastor-Satorras R., J. Stat. Mech. (2006) P05001.
% \bibitem{CastellanoPastor06}
% \Name{Castellano C. \and Pastor-Satorras R.} \REVIEW{J. Stat.
% Mech.}{}{2006}{P05001}.

\bibitem{Boyer_2003} Boyer D. and Miramontes O., Phys. Rev. E {\bf 67} (2003) 035102.
% \bibitem{Boyer_2003}
% \Name{Boyer D. \and Miramontes O.} \REVIEW{Phys. Rev. E}{67}{2003}{035102}.

\bibitem{GuimeraCartography} Guimer\'a R. and Amaral L. A. N., Nature {\bf 433} (2005) 895.
% \bibitem{GuimeraCartography}
% \Name{Guimer\'a R. \and Amaral L. A. N.}.
% \REVIEW{Nature}{433}{2005}{895}.

\bibitem{GirvanPNAS} Girvan M. and Newman M. E. J., Proc. Natl. Acad. Sci. USA {\bf 99} (2002) 7821.
% \bibitem{GirvanPNAS}
% \Name{Girvan M. \and Newman M. E. J.} \REVIEW{Proc. Natl. Acad.
% Sci. USA}{99}{2002}{7821}.

\bibitem{NewmanCommunity} Newman M. E. J. and Girvan M., Phys. Rev. E {\bf 69} (2004) 026113.
% \bibitem{NewmanCommunity}
% \Name{Newman M. E. J. \and Girvan M.} \REVIEW{Phys. Rev.
% E}{69}{2004}{026113}.

\bibitem{OverlappingCommunities}  Palla G., Der\'enyi I., Farkas I. and Vicsek T., Nature {\bf 435} (2005) 814.
% \bibitem{OverlappingCommunities}
%  \Name{Palla G., Der\'enyi I., Farkas I. \and Vicsek T.}
% \REVIEW  {Nature}{435}{2005}{814}.

\bibitem{Maslou} Maslov S. and Sneppen K., Science {\bf 296} (2002) 910.
% \bibitem{Maslou}
%    \Name{Maslov S. \and Sneppen K.}
%    \REVIEW{Science}{296}{2002}{910}.

 \end{thebibliography}
\end{document}